\documentclass{ws-ijmpa}
\usepackage{amssymb}

\begin{document}

\markboth{Tao Huang, Zuo-Hong Li, Xing-Gang Wu and Fen Zuo}
{Semileptonic $B$($B_s, B_c$) decays in the light-cone QCD sum
rules}

\catchline{}{}{}{}{}

\title{Semileptonic $B$($B_s, B_c$) decays in the light-cone QCD sum
rules}

\author{Tao Huang\footnote{Email: huangtao@mail.ihep.ac.cn}}
\address{Institute of High Energy Physics, Chinese Academy of
Sciences, Beijing 100049, China}

\author{Zuo-Hong Li}
\address{Department of Physics, Yantai University, Yantai 264005,
China}

\author{Xing-Gang Wu}
\address{Department of Physics, Chongqing University, Chongqing
400044, China}

\author{Fen Zuo}
\address{Department of Modern Physics, University of Science and
Technology of China, Hefei, Anhui 230026, China\\
Institute of High Energy Physics, Chinese Academy of Sciences,
Beijing 100049, China}

\maketitle

\begin{history}
\received{Day Month Year} \revised{Day Month Year}
\end{history}

\begin{abstract}
Semileptonic $B$($B_s, B_c$) decays are investigated systematically
in the light-cone QCD sum rules. Special emphasis is put on the LCSR
calculation on weak form factors with an adequate chiral current
correlator, which turns out to be particularly effective to control
the pollution by higher twist components of spectator mesons. The
result for each channel depends on the distribution amplitude of the
the producing meson. The leading twist distribution amplitudes of
the related heavy mesons and charmonium are worked out by a model
approach in the reasonable way. A practical scenario is suggested to
understand the behavior of weak form factors in the whole
kinematically accessible ranges. The decay widths and branching
ratios are estimated for several $B$($B_c$) decay modes of current
interest. 
\end{abstract}

\ccode{PACS numbers: 13.20.He, 12.38.Lg, 14.40.Nd, 14.40.Lb }

\maketitle

\section{Introduction}
Semileptonic $B$($B_s, B_c$) decays provide an important ground to
understand and test the standard model(SM), and perhaps, a window
into new physic beyond the SM. To test the SM with the experimental
data demands that we have the ability to precisely compute the
physical amplitudes in QCD theory. In this talk we focus on a
discussion about how to compute the related hadronic matrix elements
in the QCD light cone sum rules (LCSR's) \cite{LCSR}, which have
proved to be a powerful tool to derive the desired form factors with
a minimal number of phenomenological assumptions. Instead of the
procedure followed by conventional sum rule calculations, a LCSR
computation starts with a vacuum-meson correlator, with the related
light meson being on its mass shell, and operator product expansion
(OPE) of the correlator is performed in term of the vacuum-meson
matrix elements of certain nonlocal operators near light cone
$x^2=0$. Then one parameterizes these nonlocal matrix elements in
terms of light-cone wavefunctions of the light meson with increasing
twist. In contrast to the case of the 3-point sum rules (3PSR's),
the effective region of the momentum transfer $q^2$ can be fixed
without any extrapolation. However, it is our concern how to
effectively control the nonperturbative dynamics embedded in the
light-cone wavefunctions for enhancing the predictive power of the
LCSR method. As shown in Ref.~\cite{LCSR1}, the contributions would,
in general, be important from some of the higher twist distribution
amplitudes, especially twist-3 ones, in many situations. A pragmatic
prescription \cite{Bpi} has been put forward by us to reduce the
contamination by the higher-twists. In the improved LCSR approach,
we choose to use a certain proper chiral current operator as
interpolating fields in the related correlators, which make twist-3
wavefunctions disappear in the sum rule results. Accordingly, this
can enhance considerably one's confidence in applying LCSR's to
calculate the nonperturbative quantities.

In this paper, we would like to give the complete discussion of the
semileptonic form factors for $B$($B_s, B_c$) meson with the chiral
current correlators. A model for leading twist distribution
amplitudes is formulated in the reasonable way, for the related
heavy mesons and charmonium. Also, a practical scenario is suggested
to have an all-around understanding of the form factors in the whole
kinematical ranges.

\section{$B, B_s\to P(V)$ transition form factors}
The hadronic matrix elements for $B, B_s \to P(V)$ transitions can
be parameterized in term of the form factors in the following way:
\begin{eqnarray}
&&{<}P(p)|\bar{q}\gamma_{\mu}b|B_{(s)}(p+q){>}=f_+(q^2)(2p+q)_\mu+f_-(q^2)q_\mu,\label{eq:fBP}\\
&&{<}V(p,\eta)|\bar{q}\gamma_\mu(1-\gamma_5)b|B_{(s)}(p+q){>}{=}-i\eta^*_\mu(m_{B}{+}m_V)A_1(q^2)+i(\eta^*q)\frac{A_+(q^2)(2p+q)_\mu}{m_{B}+m_V}\nonumber\\
&&~~~~~~~~~~~~~~~~~~~~~~~~~~~~~~~~+iq_\mu(\eta^*q)\frac{A_-(q^2)}{m_{B}+m_V}+\epsilon_{\mu\alpha\beta\gamma}\eta^{*\alpha}q^\beta
p^\gamma\frac{2V(q^2)}{m_{B}+m_V},
\end{eqnarray}
where $q$ and $\eta_{\mu}$ are the momentum transfer and
polarization vector of the vector meson, respectively.

For performing a LCSR computation of the form factors for $B(B_s)\to
P$, we choose the following correlator $\Pi_\mu(p,q)$ with the
chiral current \cite{Bpi},
\begin{equation}
\Pi_\mu(p,q)=i\int
d^4xe^{iqx}{<}P(p)|T\{\bar{q}_1(x)\gamma_\mu(1+\gamma_5)b(x),\bar{b}(0)i(1+\gamma_5)q_2(0)\}|0{>}\label{eq:PCC}
\end{equation}
By contracting the $b$-quark operators to a free propagator, we get
leading twist contribution to the correlator,
\begin{eqnarray}
\Pi_\mu^{(\bar{q}q)}&=&-2m_bi\int{\frac{d^4xd^4k}{(2\pi)^4}e^{i(q-k)\cdot x}\frac{1}{k^2-m_b^2}{<}P(p)|T\bar{q}_1(x)\gamma_{\mu}\gamma_5q_2(0)|{0>}}\nonumber\\
&=&2f_Pm_bp_\mu\int_0^1{du\frac{\varphi_P(u)}{m_b^2-(up+q)^2}}+\mbox{higher
twist terms}\label{eq:Piqq}
\end{eqnarray}
where we have substituted the definition of the leading twist
distribution amplitude $\varphi_P(u)$:
\begin{equation}
{<}P(p)|T\bar{q}_1(x)\gamma_{\mu}\gamma_5q_2(0)|0{>}=-ip_{\mu}f_P\int_0^1{due^{iup\cdot
x}\varphi_P(u)}+\mbox{higher twist terms}.\label{eq:PDA}
\end{equation}
Because of the chiral nature of the selected correlator
(\ref{eq:PCC}), only the leading non-local matrix element
${<}P(p)|\bar{q}_1(x)\gamma_{\mu}\gamma_5q_2(0)|0{>}$ contributes to
the correlator, while those with leading twist-3
${<}P(p)|\bar{q}_1(x)i\gamma_5q_2(0)|0{>}$ and
${<}P(p)|\bar{q}_1(x)\sigma_{\mu\nu}\gamma_5q_2(0)|0{>}$ disappear
from the sum rule. Including the effect of the background gluon
field by writing down a full $b$-quark propagator, it is shown that
all the twist-3 terms actually vanish in the OPE of this special
correlator. Twist-4 and yet higher twist terms are found to be
numerically small, and the resulting uncertainties can always be
neglected comparing with those of the approach itself. Matching
Eq.(\ref{eq:Piqq}) to the hadronic representation of the correlator
and omitting QCD radiative corrections, we get the following simple
LCSR result for $f_+(q^2)$
\begin{equation}
f_+(q^2)=\frac{m_b^2f_P}{m_{B}^2f_{B}}e^{m_{B}^2/M^2}\int_{\Delta_P}^1{du\frac{\varphi(u)}{u}\exp{[-\frac{m_b^2-\bar{u}(q^2-um_P^2)}{uM^2}]}}+
\mbox{twist-4 terms} \label{eq:fpSR}
\end{equation}
with $\bar u=1-u$ and
\begin{equation}
\Delta_P=[\sqrt{(s_0^P-q^2-m_P^2)^2+4m_P^2(m_b^2-q^2)}-(s_0^P-q^2-m_P^2)]/(2m_P^2),\label{eq:Deltap}
\end{equation}
where $M^2$ denotes the corresponding Borel variable, and $s_0$ the
threshold parameter. When setting $M_P=0$, we get a simpler
$\Delta_P=\frac{m_b^{2}-q^2}{s^P_0-q^2}$. As a byproduct, we also
find the relation $f_-(q^2)=-f_+(q^2)$, which holds up to twist-$4$
terms. We would like to stress that the resulting form factors are
just valid for $0<q^2<m_b^2-2m_b\Lambda_{\rm{QCD}}$.

Form factor $f_+(q^2)$, as an illustrative example, has been
calculated in terms of the above expression in the cases of
$B\rightarrow\pi$ and $B_s\rightarrow K$~\cite{Bpi}. QCD radiative
corrections have been added to the twist-2 part in the $B\rightarrow
\pi$ case \cite{LCSR2}, with a negligibly small numerical impact. By
applying LCSR's to do such calculations, what is meant is that one
is assuming soft exchanges to predominate in this transition. Thus,
the semileptonic from factors in the whole kinematical range can be
determined by combing the LCSR results with those of lattice QCD,
which turn out to be available near the smallest recoil. While an
alternative choice is made, which assumes hard exchanges to dominate
at larger recoil for which the PQCD approach is applicable, a
combined use should be in order of the three different approaches
\cite{Bpi2}, to fully understand the behavior of the form factor in
the whole kinematic regions. A similar discussion on the case of
$B\to K$ form factor is presented in Ref.\cite{BK}.

\begin{figure}[pb]
\centerline{$$\epsfxsize=0.80\textwidth\epsffile{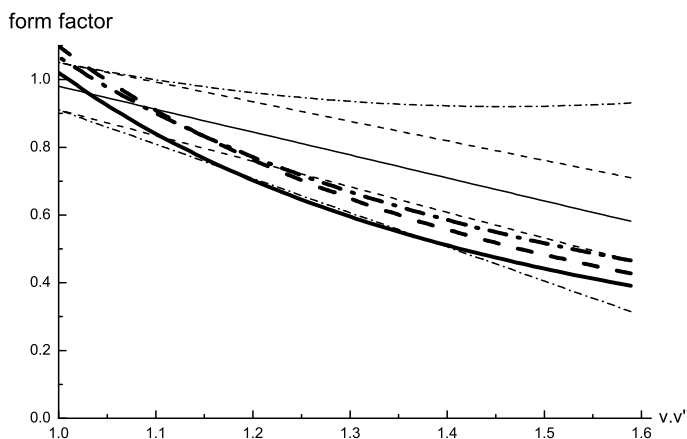}$$}
\caption{${\cal F}_{B\to D}$ as a function of the
velocity transfer. The thin lines expresses the experiment fits with
the HQET result at zero momentum transfer as input, where the solid
line represents the central values, the dashed(dash-dotted) lines
give the bounds from the linear(quadratic) fits. The thick lines
correspond to our results, with the solid, dashed and dash-dotted
lines for model {I}{I}{I}, {I}{I} and I
respectively.}\label{fig:F(y)}
\end{figure}

It is straightforward to extend our discussion to the $B\to D$ case.
In Ref.\cite{BD}, the $B\rightarrow D$ form factor is calculated
using three different models for the $D$ meson distribution
amplitude. In order to compare with the results of heavy quark
effective theory (HQET), the form factor has been redefined
according to
\begin{equation}
f_+(q^2)=\frac{m_B+m_D}{2\sqrt{m_B~m_D}}{\cal F}_{B\to D}(v\cdot
v'=\frac{m_B^2+m_D^2-q^2}{2m_B~m_D}).\label{eq:f12}
\end{equation}
The numerical results, which correspond to the region $0\le q^2 \le
(m_B-m_D)^2$, are shown in Fig.\ref{fig:F(y)}. At zero recoil ${\cal
F}_{B\to D}^{LC}(1)=1.02$ (using model {I}{I}{I} in Ref.~\cite{BD}),
which is in a good agreement with the evaluation obtained using the
heavy quark symmetry: ${\cal F}_{B\to D}(1)=0.98\pm 0.07$
\cite{HQET}. In the larger recoil region $1.35 < v\cdot v' < 1.59$,
the yielded results are consistent with those of pQCD \cite{PQCD}.
Therefore, a full understanding of the dynamics involved in $ B\to D
$ transition may be obtained by combining the three different
approaches~---~HQET, LCSR's and pQCD, which are complementary to
each other.

For $B(B_s)\to V$ modes, we choose the following correlator as our
starting point:
\begin{eqnarray}
\Pi_\mu(p,q)&=&-i\int d^4xe^{iq\cdot x}{<}V(p,\eta)|T\{\bar{q}_1(x)\gamma_\mu(1-\gamma_5)b(x),\bar{b}_1(0)(1+\gamma_5)q_2(0)\}|0{>}\nonumber\\
            &=&\Gamma^1\eta^*_\mu-\Gamma^+(\eta^*q)(2p+q)_\mu-\Gamma^-(\eta^*q)q_\mu+i\Gamma^V\varepsilon_{\mu\alpha\beta\gamma}\eta^{*\alpha}q^{\beta}p^{\gamma}.
\end{eqnarray}
Also we take the standard definition of the twist-$2$ and twist-$3$
distribution amplitudes of the vector meson:
\begin{eqnarray}
&&{<}V(p)|\bar
q_{1\beta}(x)q_{2\alpha}(0)|0{>}\nonumber=\frac{1}{4}\int^1_0due^{iup\cdot
x}
\{f_Vm_V[\eta\!\!\!/^*g^{(v)}_\perp(u)+p\!\!\!/\frac{(\eta^*x)}{(px)}(\phi_\parallel(u)-g^{(v)}_\perp(u))]\nonumber\\
&&~~~~-if^T_V\sigma_{\mu\nu}\eta^{*\mu}p^\nu\phi_\perp(u)+
\frac{m_V}{4}(f_V-f^T_V\frac{m_1+m_2}{m_V})\epsilon^\mu_{\nu\alpha\beta}
\gamma_\mu\gamma_5\eta^{*\nu} p^\alpha
x^{\beta}g^{(a)}_\perp(u)\}_{\alpha\beta},\label{eq:VDA}
\end{eqnarray}
Actually there are also two other twist-3 distribution
amplitudes~\cite{Ballrho1},
$h_\parallel^{(s)}(u),h_\parallel^{(t)}(u)$, which are related to
the square of light meson mass, and therefore can be neglected in
comparison with those written above. We also neglected twist-4 and
higher terms, and the whole three-particle contributions. Following
the standard procedure, one obtains:
\begin{eqnarray}
&&A_1(q^2)=\frac{f^T_Vm_b}{f_{B}m_B^2(m_B+m_V)}e^{m_B^2/M^2_V}\nonumber\\
&&~~~~~~~~~~~~\times \int^1_{\Delta_V}\frac{du}{u}\exp{\left[-\frac{m_b^2-\bar{u}(q^2-um_V^2)}{uM_V^2}\right]}\frac{m_b^2-q^2+u^2m_V^2}{u}\phi_\perp(u),\\
&&A_+(q^2)=\frac{f^T_Vm_b(m_{B}+m_V)}{f_Bm_B^2}e^{m_B^2/M^2_V}\nonumber\\
&&~~~~~~~~~~~~\times \int^1_{\Delta_V}\frac{du}{u}\exp{\left[-\frac{m_b^2-\bar{u}(q^2-um_V^2)}{uM_V^2}\right]}\phi_\perp(u),\\
&&A_-(q^2)=-A_+(q^2),\\
&&V(q^2)~~=A_+(q^2)
\end{eqnarray}
with
\begin{equation}
\Delta_V=[\sqrt{(s_0^V-q^2-m_V^2)^2+4m_V^2(m_1^2-q^2)}-(s_0^V-q^2-m_V^2)]/(2m_V^2),\label{eq:delta2}
\end{equation}

The same technic can be used to discuss the rare decay $B\to
V\gamma$. The electromagnetic penguin operator dominates in this
transition, and relevant decay amplitude reads
\begin{eqnarray}
A(B \to V\gamma)=Cm_b\epsilon^{\mu}<V(p,\eta)|\bar q
\sigma_{\mu\nu}(1+\gamma_5)q^{\nu}b|B(p+q)>,
\end{eqnarray}
where $\epsilon_{\mu}$ and $q$ are the emitted photon polarization
vector and momentum, respectively. The constant $C$ depends on the
product of CKM matrix elements $V_{tb}V_{td}^*$ and the
corresponding Wilson coefficient $C_7$. The hadronic matrix element
may be parameterized in terms of the form factor $T(q^2=0)$,
\begin{equation}
<V(p,\eta)|\bar{q}\sigma_{\mu\nu}(1+\gamma_5)q^{\nu}b|B(p+q)>=2(i\epsilon_{\mu\nu\alpha\beta}
                 \eta^{*\nu}q^{\alpha}p^{\beta}
                 +p\cdot{q}\eta^{*}_{\mu}
                 -q\cdot{\eta^{*}p_{\mu}})T(0).
\end{equation}
A correlator, which is suitable for our purpose, is
\begin{eqnarray}
F_{\mu}(p,q) &&=i\int d^{4}xe^{iq\cdot x}<V(p,\eta)|T\bar{q_1}(x)
\sigma_{\mu\nu}
(1+\gamma_5) q^{\nu}b(x),\bar{b}(0) i(1+\gamma_5)q_2(0)|0>\nonumber\\
&&=\left[2i\epsilon_{
\mu\nu\alpha\beta}\eta^{\ast\nu}q^{\alpha}p^{\beta}+2p\cdot
q\eta_{\mu}^{\ast}-2q\cdot \eta^{\ast}
p_{\mu}\right]F\left[(p+q)^2\right].
\end{eqnarray}

According to the definition in Eq.(\ref{eq:VDA}) we can parameterize
the nonlocal matrix element
$<V(p,\eta)|\bar{q_1}(x)\sigma_{\mu\nu}q^{\nu}(1+\gamma_5)q_2(0)|0)>$
using the the leading twist-2 wavefunction $\phi_{\perp} (u)$ as the
following,
\begin{eqnarray}
<V(p,\eta)|\bar{q_1}(x)\sigma_{\mu\nu}q^{\nu}(1+\gamma_5)q_2(0)|0>
&&=i\left[(q\cdot\eta^{*})p_{\mu}-p\cdot{q}\eta^{*}_{\mu}-i\epsilon_{
\mu\nu\alpha\beta}\eta^{\nu}q^{\alpha}p^{\beta}\right]f^{V}_{\perp}\nonumber\\
&&\times\int\limits_{0}^{1}due^{iup\cdot x}\phi_{\perp}(u).
\end{eqnarray}
Then we get the final LCSR for the form factor $T(0)$
\begin{eqnarray}
T(q^2=0)=\frac{m_b^2f^T_V}{m_{B}^2f_{B}}e^{m_{B}^2/M_V^2}
\int_{\Delta_V^0}^1{du\frac{\phi_{\perp}(u)}{u}\exp{[-\frac{m_b^2+u\bar{u}m_V^2)}{uM_V^2}]}},
\end{eqnarray}
where $\Delta_V^0=\Delta_V|_{q^2=0}$. Notice that
\begin{equation}
T(q^2=0)=\frac{m_b}{(m_B+m_V)}A_+(q^2=0)
\end{equation}

With this sum rule, both $B\to K^*\gamma$ and $B\to
(\rho,\omega)\gamma$ have been discussed in Ref.~\cite{BK*,Brho}. In
Ref.~\cite{Brho}, the form factor and branching ratio for $B\to
(\rho,\omega)\gamma$ are achieved using two different models for the
$\rho$ meson distribution amplitude:
\begin{equation}
T^{B\to\rho^{\pm}}(0)=0.335\pm0.050, \
Br(B\to\rho^{\pm}\gamma)=(2.71\pm1.00) \times10^{-6} \nonumber
\end{equation}
for the model introduced by Ball and Braun in Ref.~\cite{Ballrho2},
and
\begin{equation}
T^{B\to\rho^{\pm}}(0)=0.272\pm0.029,\
Br(B\to\rho^{\pm}\gamma)=(1.79\pm0.61)
\times10^{-6} \nonumber\\
\end{equation}
for the model presented by Bakulev and Mikhailo in
Ref.~\cite{Bakulev}. Comparing the LCSR predictions with the recent
experimental result \cite{Brho2}
\begin{equation}
Br(B\to\rho^{\pm}\gamma)=(1.32^{+0.34}_{-0.31}(\mbox{stat})^{+0.10}_{-0.09}(\mbox{syst}))
\times10^{-6}, \nonumber
\end{equation}
we can determine which model can better describe the $\rho$ meson.

\section{$B_{c}\to P(V)$ transition form factors}
$B_c\to P(V)$ situations may be discussed in parallel by a
corresponding displacement of inputs. We proceed to cope with the
semileptonic transitions induced, respectively, by $b\to u, c $ and
$c \to d, s$. However, since the related distribution amplitudes are
not available from a QCD calculation, we give a model description
for them, which is based on the harmonic oscillator
potential~\cite{Bc}.
\begin{table}[h]
\tbl{The values of the form factors at $q^2=0$ in comparison with
the estimates in the three points sum rule (3PSR) (with the Coloumb
corrections included) and in the quark model (QM).}
{\begin{tabular}{@{}cc|cc|c c c c@{}} \toprule
Mode&&$f_+(0)$&$f_-(0)$&$A_1(0)$&$A_+(0)$&$A_-(0)$&$V(0)$\\ \colrule
                    &   This work      &   0.87  &  -0.87  & 0.75 & 1.69 & -1.69 & 1.69 \\
$B_c\to \bar cc[1S]$&   3PSR~\cite{3PSR2}&   0.66  &  -0.36  & 0.63 & 0.69 & -1.13 & 1.03 \\
                    &   QM~\cite{RQM}    &   0.76  &  -0.38  & 0.68 & 0.66 & -1.13 & 0.96
                    \\\hline
                    & This work      & 1.02 &  -1.02  & 1.01 & 9.04 & -9.04 & 9.04 \\
$B_c\to B_s^{(*)}$  & 3PSR~\cite{3PSR2}& 1.3  &  -5.8   & 0.69 &-2.34 & -21.1 & 12.9 \\
                    & QM~\cite{RQM}    &-0.61 &   1.83  &-0.33 & 0.40 &  10.4 & 3.25 \\
                    \hline
                    & This work        & 0.90 &  -0.90  & 0.90 &  7.9 &  -7.9 &  7.9 \\
$B_c\to B^{(*)}$    & 3PSR~\cite{3PSR2}& 1.27 &  -7.3   & 0.84 & -4.06&  -29.0& 15.7 \\
                    & QM~\cite{RQM}    &-0.58 &   2.14  &-0.27 &  0.60&   10.8& 3.27 \\
                    \hline
                    & This work      & 0.35 &  -0.35  & 0.32 & 0.57 & -0.57 & 0.57 \\
$B_c\to D^{(*)}$    & 3PSR~\cite{3PSR2}& 0.32 &  -0.34  & 0.43 & 0.51 & -0.83 & 1.66 \\
                    & QM~\cite{RQM}    & 0.69 &  -0.64  & 0.56 & 0.64 & -1.17 & 0.98 \\ \botrule
\end{tabular}  \label{tab:ff}}
\end{table}

\begin{table}[h]
\tbl{Branching ratios (in \%) of simileptonic $B_c$ decays into
ground state charmonium states, and into ground charm and bottom
meson states, in comparison with the result of 3PSR, QM, and the
approach of the Bethe-Salpeter equation. For the lifetime of the
$B_c$ we take $\tau(B_c)=0.45\rm{ps}$.}
{\begin{tabular}{c@{}c@{}|@{}c@{}|@{}c@{}|@{}c@{}|@{}c@{}|@{}c@{}|@{}c@{}|@{}c@{}|@{}c@{}|@{}c@{}|@{}c@{}|@{}c@{}}
\toprule
  Mode           &$~\eta_ce\nu$&$~\eta_c\tau\nu$&$J/\psi
  e\nu$&$J/\psi\tau\nu$&$~~De\nu~$&$~~D\tau\nu~$&$~D^*e\nu~$&$~D^*\tau\nu~$&$~Be\nu$&$~B^*e\nu$&$~B_se\nu$&$~B^*_se\nu$\\
  \colrule
This work               & 1.64 & 0.49 & 2.37 & 0.65 &  0.020 &  0.015 & 0.035 & 0.020 & 0.21 & 0.32 & 3.03 & 4.63
\\\hline
3PSR~\cite{3PSR2}  & 0.75 & 0.23 & 1.9  & 0.48 & 0.004  &  0.002 & 0.018 & 0.008 & 0.34 & 0.58 & 4.03 & 5.06 \\
\hline
QM~\cite{RQM}      & 0.98 & 0.27 & 2.30 & 0.59 & 0.018  &  0.0094& 0.034 & 0.019 & 0.15 & 0.16 & 2.00 & 2.6\\
\hline
BSE \cite{BSE}        & 0.97 & ---  &
2.30&---&0.006&---&0.018&---&0.16 &0.23 &1.82&3.01\\ \botrule
\end{tabular} \label{tab:BR}}
\end{table}

Our calculations, as in the $B$ meson case, are limited to the
regions where the OPE goes effectively, namely,
$0<q^2<m_b^2-2m_b\Lambda_{\rm{QCD}}\simeq15~\mbox{GeV}^2 $for
$b$-quark decays and
$0<q^2<m_c^2-2m_c\Lambda_{\rm{QCD}}\simeq0.4~\mbox{GeV}^2 $ for
$c$-quark decays. The numerical results for the form factors at
$q^2=0$ are collected in Tab.1. It turns out that the calculated
form factors can be parameterized excellently as
\begin{equation}
F_i(q^2)=\frac{F_i(0)}{1-a_iq^2/m_{B_c}^2+b_i(q^2/m_{B_c}^2)^2}\label{eq:para}.
\end{equation}
Extrapolating the LCSR results with this parametrization, we can
estimate the branching ratios of the simileptonic $B_c$ decays. The
results are shown in Tab.2 together with those of other approaches,
where we have input the following CKM-matrix elements:
$V_{cb}=0.0413,~V_{ub}=0.0037,~V_{cs}=0.974,~V_{cd}=0.224$. For the
$b$-quark decay modes, our results for the branching ratios are much
larger than the corresponding those of 3PSR's. In these decays, the
kinematical region is rather large and the main contributions to
branching ratios should come from QCD dynamics at smaller recoil.
The numerical values of the form factors always increase much faster
with $q^2$ in the LCSR approach than in the simple pole
approximation required in the 3PSR analysis, which accounts for the
numerical discrepancy between the two approaches. While in the
$c$-quark decays, where the kinematical region is narrow enough, the
results of the two approaches are comparable to each other.

\section{summary}
In this paper we derive LCSR's for the form factors for semileptonic
$B$($B_s, B_c$) decays of current interest, which depend mainly on
the leading twist distribution amplitudes of the producing mesons by
using suitable chiral currents.

For better understanding the behavior of these form factors in the
whole kinematical range, a combined use is necessary of several
different approaches which could be complementary to each other, for
different kinematical regions. For example, in the $B\to D$ case one
can use three different methods (LCSR, pQCD and HQET) for different
kinematical regions to understand it in the whole kinematical range.

Based on the models with the harmonic oscillator potential for the
light-cone wave functions, we calculate the form factors for several
important semileptonic $B_c$ decay modes in the their kinematical
regions allowed by LCSR calculation. Extrapolating the LCSR results
to the large $q^2$ regions, we derive the decay widths and branching
ratios.

\begin{center}
{\bf ACKNOWLEDGEMENTS}
\end{center}

This work was supported in part by the Natural Science Foundation of
China (NSFC).


\begin{thebibliography}{99}

\bibitem{LCSR}V. L. Chernyak, and I. R. Zhitnitsky, {\it Nucl. Phys} {\bf
B 345}, 137 (1990); I. I. Balitsky, V. M. Braun, and A. V.
Kolesnichenko {\it Nucl. Phys. } {\bf B 312}, 509 (1989).

\bibitem{LCSR1}V. M. Belyaev, A. Khodjamirian and R. R\"{u}ckl, {\it Z. Phys.} {\bf C 60}, 349 (1993); V. M. Belyaev, V. M. Braun, A. Khodjamirian, R. R\"{u}ckl, {\it Phys. Rev. } {\bf D 51}, 6177
(1995); A. Khodjamirian, R. R\"{u}ckl, S. Weinzierl, and O.
Yakovlev, {\it Phys. Lett} {\bf B 410}, 275 (1997); E. Bagan, P.
Ball, and V. M. Braun, {\it Phys. Lett} {\bf B 417}, 154 (1998).

\bibitem{Bpi}T. Huang, Z. H. Li, and X. Y. Wu,
{\it Phys. Rev. } {\bf D 63}, 094991 (2001); Z. H. Li, F. Y. Liang,
X. Y. Wu and T. Huang, {\it Phys. Rev. } {\bf D64}, 057901 (2001).

\bibitem{LCSR2}Z. G. Wang, M. Z. Zhou, and T. Huang, {\it Phys. Rev. } {\bf D 67}, 094006 (2003).

\bibitem{Bpi2}T. Huang, and X-G. Wu,
{\it Phys. Rev. } {\bf D 71}, 034018 (2005).

\bibitem{BK} X-G. Wu, T. Huang and Z. Y. Fang, {\it Eur. Phys. J.} {\bf C52},
561(2007); arXiv: 07120237.

\bibitem{BD}F. Zuo, Z. H. Li and T. Huang, {\it Phys. Lett. } {\bf B 641}, 177 (2006); F. Zuo, and T. Huang, {\it Chin. Phys. Lett.} {\bf 24},
61 (2007).

\bibitem{HQET}I. Caprini, L. Lellouch, and M. Neubert,
{\it Nucl. Phys. } {\bf B 530}, 153 (1998).

\bibitem{PQCD}T. Kurimoto, H. N. Li and A. I. Sanda,
{\it Phys. Rev. } {\bf D 67}, 054028 (2003).

\bibitem{Ballrho1}P. Ball and V. M. Braun, {\it Phys. Rev. } {\bf
D58}, 094016 (1998).

\bibitem{BK*}T. Huang and Z. H. Li, {\it Phys. Rev.} {\bf D 57}, 1993 (1998).

\bibitem{Brho}T. Huang, Z. H. Li, and H. D. Zhang, {\it J. Phys. } {\bf G 25}, 1179 (1999).

\bibitem{Ballrho2}P. Ball and V. M. Braun, {\it Phys. Rev. } {\bf D 54}, 2182 (1996).

\bibitem{Bakulev}A. P. Bakulev and S. V. Mikhailov, {\it Phys. Lett. } {\bf B 436}, 351 (1998).

\bibitem{Brho2}D. Mohapatra et al. (Belle Collaboration) {\it Phys. Rev. Lett. } {\bf D 96}, 221601 (2006).

\bibitem{Bc}T. Huang, and F. Zuo, {\it Euro. Phys. J. } {\bf C 51}, 833 (2007).

\bibitem{3PSR2}V. V. Kiselev, A. K. Likhoded and A. I. Onishchenko, {\it
Nucl. Phys. } {\bf B 569}, 473 (2000); {\bf B 585}, 353 (2000); V.
V. Kiselev, arXiv:hep-ph/0211021.


\bibitem{RQM}M. A. Ivanov, J. G. K\"{o}rner, P. Santorelli,
{\it Phys. Rev. } {\bf D63}, 074010 (2001).

\bibitem{BSE}C. H. Chang and Y. Q. Chen,
{\it Phys. Rev. } {\bf D49}, 3399 (1994).



\end{thebibliography}
\end{document}